\documentclass[preprint,aps]{revtex4}
\usepackage{graphicx}
\usepackage{latexsym}

\begin{document}

\title{Optimization by thermal cycling}

\author{A.~M\"obius}

\address{Leibniz Institute for Solid State and Materials Research Dresden,
  \linebreak  PF 27\,01\,16, D-01171 Dresden, Germany\\
  E-mail: a.moebius@ifw-dresden.de}

\author{K.H.~Hoffmann}

\address{TU Chemnitz, Institute of Physics, D-09107 Chemnitz, Germany\\
  E-mail: hoffmann@physik.tu-chemnitz.de}

\author{C.~Sch\"on}
\address{Max Planck Institute for Solid State Research, D-70569
  Stuttgart, Germany\\
  E-mail: schoen@fkf.mpg.de\\}

\begin{abstract}
Thermal cycling is an heuristic optimization algorithm which consists of 
cyclically heating and quenching by Metropolis and local search 
procedures, respectively, where the amplitude slowly decreases. In recent 
years, it has been successfully applied to two combinatorial optimization 
tasks, the traveling salesman problem and the search for low-energy 
states of the Coulomb glass. In these cases, the algorithm is far more 
efficient than usual simulated annealing. In its original form the 
algorithm was designed only for the case of discrete variables. 
Its basic ideas are applicable also to a problem with continuous 
variables, the search for low-energy states of Lennard-Jones clusters.
\end{abstract}

\maketitle

\section{Introduction}

Optimization problems with large numbers of local minima occur in 
many fields of physics, engineering, and economics. They are closely 
related to statistical physics, see e.g.\  Ref.\ [1]. In the case of 
discrete variables, such problems often arise from combinatorial 
optimization tasks. Many of them are difficult to solve since they are 
NP-hard, i.e., there is no algorithm known which finds the exact 
solution with an effort proportional to any power of the problem size. 
One of the most popular such tasks is the traveling salesman problem: 
how to find the shortest roundtrip through a given set of cities 
\cite{John.McGe}.

Many combinatorial optimization problems are of considerable practical
importance. Thus, algorithms are needed which yield good approximations 
of the exact solution within a reasonable computing time, and which 
require only a modest effort in programming. Various deterministic and 
probabilistic approaches, so-called search heuristics, have been 
proposed to construct such approximation algorithms. A considerable part 
of them borrows ideas from physics and biology. Thus simulated annealing
\cite{Kirk.etal} and relatives such as threshold accepting as well as 
various genetic algorithms \cite{hol75:ana} have successfully been 
applied to many problems. Particularly effective seem to be genetic 
algorithms in which the individuals are local minima 
\cite{Brad,Merz.Frei}. For recent physically motivated heuristic 
approaches we refer to thermal cycling \cite{thercycl}, optimization by 
renormalization \cite{renormopt}, and extremal optimization 
\cite{extropt}. For problems with continuous variables, approaches which 
combine Monte-Carlo procedures for global search with deterministic 
local search by standard numerical methods, for example the 
basin-hopping algorithm, have proved to be particularly efficient 
\cite{bashop,bashopjump}. They can be considered as relatives of the 
genetic local search approaches for the case of discrete variables. Here 
we focus on the thermal cycling algorithm and illuminate the reasons for 
its efficiency.

\section{Thermal cycling algorithm}
Simulated annealing \cite{Kirk.etal} can be understood as a random
journey of the sample (i.e.\ the approximate solution) through a
hilly landscape formed by the states of its configuration space. The
altitude, in the sense of a potential energy, corresponds to the
quantity to be optimized. In the course of the journey, the altitude
region accessible with a certain probability within a given number of
steps shrinks gradually due to the decrease of the temperature in the
Metropolis simulation involved. The accessible area, i.e., the 
corresponding configuration space volume, thus shrinks until the sample 
gets trapped in one of the local minima.

Deep valleys attract the sample mainly by their area. However, it is
tempting to make use of their depth. For that, we substitute the
slow cooling down by a cyclic process: First, starting from the lowest
state obtained so far, we randomly deposit energy into the sample by
means of a Metropolis process with a certain temperature $T$, which is
terminated, however, after a small number of steps. This part is
referred to as heating. Then we quench the sample by means of a local
search algorithm. Heating and quenching are cyclically repeated where 
the amount of energy deposited in a cycle decreases gradually, see 
Fig.\ 1. This process continues until, within a `reasonable' CPU time, 
no further improvement can be found. 

It is an essential feature of the thermal cycling algorithm that two 
contradicting demands are met in heating: the gains of the previous 
cycles have to be retained, but the modifications must be sufficiently 
large, so that another valley can be reached. Thus the heating process 
has to be terminated in an early stage of the equilibration. An 
effective method is to stop it after a fixed number of successful 
Metropolis steps.

The efficiency of the proposed algorithm depends to a large extent on 
the move class considered in the local search procedure.  For discrete 
optimization problems, it is a great advantage of our approach that far 
more complex moves can be taken into account than in simulated 
annealing so that the number of local minima is considerably reduced. 
The local search concerning complex moves can be enormously sped up 
by use of branch-and-bound type algorithms. Their basic idea is to 
construct new trial states following a decision tree: At each branching 
point, a lower bound of the energy of the trial state is calculated. The 
search within the current branch is terminated as soon as this bound 
exceeds the energy of the initial state.

The basic thermal cycling procedure can be easily accelerated in three 
ways: \linebreak (i) partition of the computational effort into several 
search processes in order to minimize the failure risk \cite{Hube.etal}, 
(ii) restricting the moves in heating to the `sensible sample regions' 
by analyzing previous cycles, or by comparing with samples considered in 
parallel, and (iii) combining parts of different states 
\cite{thercycl,itertran}.

\section{Applications}
The thermal cycling algorithm was first tested on the traveling salesman
problem \cite{thercycl}. For that, we considered problems of various 
size from the TSPLIB95 \cite{TSPLIB95} for which the exact solutions, or 
at least related bounds, are known. Fig.\ 2 gives a comparison of 
thermal cycling data with results from simulated annealing and from 
repeated local searches starting from random states. For a meaningful
characterization of the algorithms, it relates mean deviations from the 
optimum tour length to the CPU-time effort for various parameter values.
The diagram includes data for two move classes: (a) cutting a roundtrip
twice, reversing the direction of one of its parts, and connecting the 
parts then again, or shifting a city from one to another position in the 
roundtrip; (b) same as (a) and additionally rearrangements by up to four 
simultaneous cuts as well as Lin-Kernighan realignments \cite{Lin.Kern}. 

Fig.\ 2 shows that, for the traveling salesman problem, thermal cycling 
is clearly superior to simulated annealing, already if the same move 
class is considered in both procedures -- the simulated annealing code 
had been carefully tuned too --.  However, when taking advantage of the 
possibility to incorporate more complex moves, thermal cycling beats 
simulated annealing by orders of magnitude in CPU time. Applied to an 
archive of samples instead of to a single one, it can compete with 
leading genetic local search algorithms \cite{thercycl,itertran}.
 
For several years, we have used thermal cycling as standard approach
in numerical investigations of the Coulomb glass, which is basically
an Ising model with long-range interactions. Also in this case, thermal 
cycling proved to be a very efficient tool. Fig.\ 3 presents data from 
an investigation comparing several \linebreak algorithms 
\cite{Coulombglass}. In simulated annealing, we could efficiently treat 
only particle exchange with a reservoir and one-particle hops inside the 
sample, that is occupation changes of one or two sites. However, in the 
deterministic local search, the simultaneous occupation modification of 
up to four sites could be considered by means of branch-and-bound 
approaches. Therefore, the corresponding multistart local search yields 
significantly better results than simulated annealing. Thermal cycling 
of the low-energy states proves to be still far more efficient than the 
local search repeatedly starting from random states.

It is tempting to apply the thermal cycling approach also to problems 
with continuous variables. Thus we have considered Lennard-Jones cluster 
of various size because the energy landscapes of this system are known 
to have large numbers of local minima. The heating consisted of 
simulateneously shifting all atoms by small distances a few times 
according to a thermal rejection rule, and the quench combined the 
Powell algorithm with a systematic consideration of symmetry positions. 
The ground states of several clusters of up to 150 atoms could be 
reproduced within `reasonable' CPU times. Further related investigations 
should be promising.

\begin{figure}[hp] 
\includegraphics[width=0.70\linewidth]{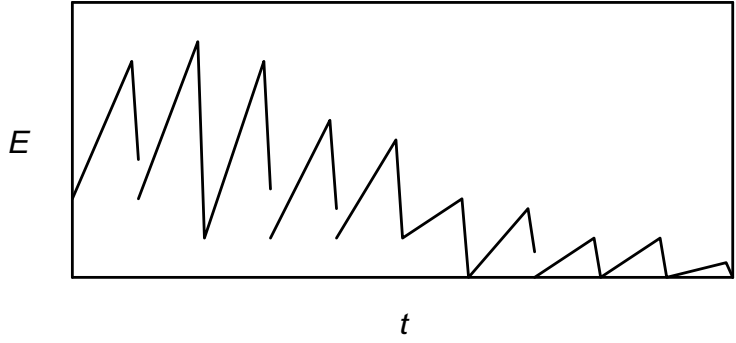}
\caption{Time dependence of the energy $E$ (quantity to be optimized) of 
the sample currently treated in the cyclic process. Gaps in the curve 
refer to cycles where the final state has a higher energy than the 
initial state, so that the latter is used as initial state of the next 
cycle too.}
\end{figure}

\begin{figure}[hp] 
\includegraphics[width=0.70\linewidth]{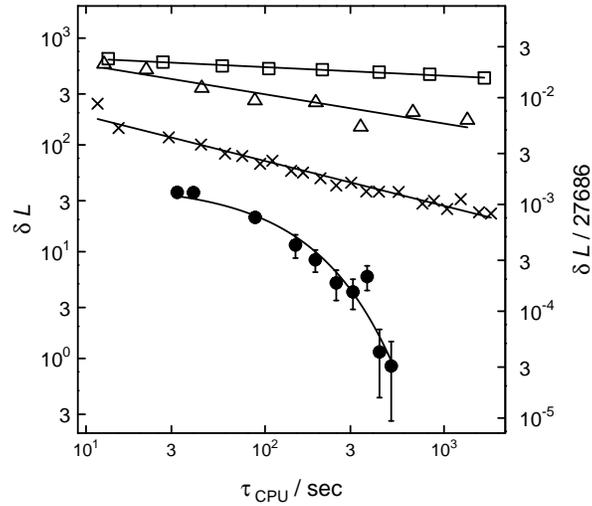}
\caption{Relation between CPU time, $\tau_{\rm {CPU}}$ (in seconds for 
one PA8000 180 MHz processor of an HP K460), and deviation,
$\delta L = L_{\rm {mean}} - 27686$, of the obtained mean approximate
solution from the optimum tour length for the Padberg-Rinaldi 532 city
problem {\protect \cite{thercycl}}. $\Box$: repeated quench to stability 
with respect to move class (a) defined in the text; $\bigtriangleup$: 
simulated annealing; $\times$ and  $\bullet$: thermal cycling with 
ensembles of various size, and local search concerning move classes (a) 
and (b), respectively. In all cases, averages were taken from 20 runs. 
Errors ($1\sigma$-region) are presented if they exceed the symbol size. 
The lines are guides to the eye only.}
\end{figure}

\begin{figure}[hp] 
\includegraphics[width=0.70\linewidth]{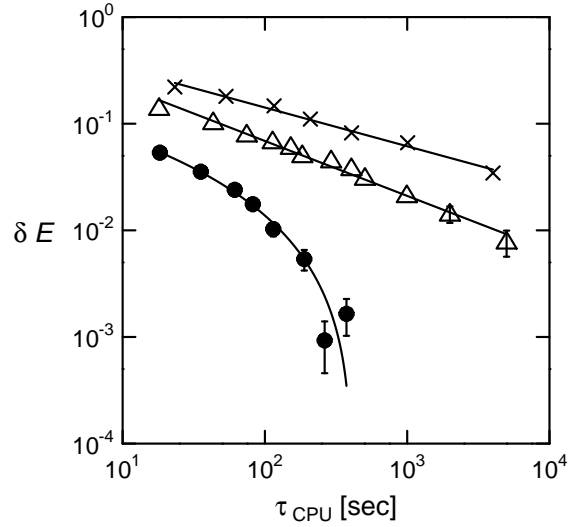}
\caption{
Mean deviation of the energy of the lowest state found from the ground 
state energy, $\delta E = E_{\rm mean}-E_{\rm ground\ state}$, related 
to the CPU time $\tau_{\rm CPU}$ (180 MHz PA8000 processor of HP K460) 
for one  realization of the three-dimensional Coulomb glass lattice 
model with 1000 sites, half filling, and medium disorder strength
{\protect \cite{Coulombglass}}. $\times $: simulated annealing; 
$\triangle $: multistart local search considering simultaneous 
occupation changes of up to four sites; $\bullet $: thermal cycling. For 
simulated annealing and multistart local search, averages
were taken from 20 runs, for thermal cycling from 100 runs. In thermal
cycling, the ground state was always found within 500 seconds.}
\end{figure}

\end{document}